# A Representation of Quantum Measurement in Order-Unit Spaces

Gerd Niestegge

Zillertalstrasse 39, 81373 Muenchen, Germany
gerd.niestegge@web.de

*Abstract.* A certain generalization of the mathematical formalism of quantum mechanics beyond operator algebras is considered. The approach is based on the concept of conditional probability and the interpretation of the Lüders - von Neumann quantum measurement as a probability conditionalization rule. A major result shows that the operator algebras must be replaced by order-unit spaces with some specific properties in the generalized approach, and it is analyzed under which conditions these order-unit spaces become Jordan algebras. An application of this result provides a characterization of the projection lattices in operator algebras.

## 1. Introduction

Despite of many efforts, quantum mechanics and relativity theory have been resisting their unification for almost a century. Either both theories are universally valid and the right way how to unify them has still not been found, or at least one of the two theories needs to be extended to achieve their unification. Just as classical probability theory was not general enough to cover the probabilities occurring in quantum mechanics, the current mathematical formalism of quantum mechanics may still not be general enough for the unification, or this may hold for relativity theory or for both theories. The present paper deals with the first case - i.e., with a potential generalization of the mathematical formalism of quantum mechanics.

The starting point is a certain axiomatic approach to this formalism, developed by the author in some recent papers [21-23], which is based on the concept of conditional probabilities and the interpretation of the Lüders - von Neumann quantum measurement as a probability conditionalization rule. This approach leads to the standard model of quantum mechanics with the Hilbert space over the complex numbers. When the last one among the axioms is dropped, it leads to Jordan algebras, which are only a little bit more general than the standard Hilbert space model since most of them - but not all - have representations as operator algebras on a complex Hilbert space. The approach provides further opportunities for potential generalizations just by dropping some more axioms and keeping only those that describe the basic properties of the conditional probabilities.

In the paper, it is shown how this results in a certain new mathematical structure - an order-unit space with some specific additional properties. It still features some of the well-known properties of Jordan operator algebras (e.g., existence and uniqueness of the conditional probabilities), but not all (e.g., the spectral theorem does not hold anymore for all elements of the order-unit space). Conditions under which the order-unit space becomes a Jordan operator algebra are studied. An application of the main results of the paper provides a characterization of the projection lattices in Jordan operator algebras, which modifies an earlier result by Bunce and Wright [7].

Order-unit spaces were introduced by Kadison [17] and play an important role in the study of convex sets [1,2,6] as well as of the order structure of operator algebras [14]. Alfsen and Shultz [2] used them in their approach to a non-commutative spectral theory. Bunce, Wright [7] and Pulmannová [25] studied the relation of Alfsen's and Shultz's results to quantum logic. Edwards and Rüttimann [12] investigated the conditional probabilities assuming the Alfsen-Shultz property (i.e., every exposed face of the state space is projective). All these results are based on the notion of P-projections introduced by Alfsen and Shultz. A similar, but more general type of projections will play a central role in the present paper; they become identical with the P-projections only in specific situations.

The next section gives an overview of the basic notions from the author's recent papers [21-23] as far as needed in the present paper. A first major result is the derivation of the order-unit space and its specific properties from these basic notions in section 3. Observables and the relationship with Alfsen and Shultz's spectral duality are studied in section 4. Further results in the last section concern the conditions under which the order unit spaces become Jordan algebras and the characterization of the projection lattices in operator algebras.

## 2. Events, states, and conditional probabilities

Our model of events (or quantum logic [7,25,28,29]) shall be a mathematical structure which is as simple as possible, but has enough structure for the consideration of states. This requires an orthogonality relation and a sum operation for orthogonal events. The precise axioms for the system $E$ of events were presented in [21] and look as follows.

$E$ is a set with distinguished elements 0 and $\mathbb{I}$, an orthogonality relation $\perp$ and a partial binary operation + such that the following conditions hold for $e,f,g \in E$:

(OS1)  *If $e \perp f$, then $f \perp e$; i.e., the relation $\perp$ is symmetric.*
(OS2)  *$e+f$ is defined for $e \perp f$, and then $e+f=f+e$; i.e., the sum operation is commutative.*
(OS3)  *If $g \perp e$, $g \perp f$, and $e \perp f$, then $g \perp e+f$, $f \perp g+e$ and $g+(e+f)=(g+e)+f$; i.e., the sum operation is associative.*
(OS4)  *$0 \perp e$ and $e+0=e$ for all $e \in E$.*
(OS5)  *For every $e \in E$, there exists a unique $e' \in E$ such that $e \perp e'$ and $e+e'=\mathbb{I}$.*
(OS6)  *There exists $d \in E$ such that $e \perp d$ and $e+d=f$ if and only if $e \perp f'$.*

Then $0'=\mathbb{I}$ and $e''=e$ for $e \in E$. A further relation $\prec$ is defined on $E$ via $e \prec f$ iff $e \perp f'$. This relation will be needed for the definition of the conditional probabilities, but note that the above axioms do not imply that it is an order relation. We call $E$ orthogonally $\sigma$-complete if the sum exists for any countable orthogonal subset of $E$, and we call $E$ orthogonally complete if the sum exists for any orthogonal subset of $E$.

A state is a map $\mu: E \to [0,1]$ such that $\mu(\mathbb{I})=1$ and $\mu(e+f) = \mu(e) + \mu(f)$ for orthogonal pairs $e$ and $f$ in $E$. Then $\mu(0)=0$ and $\mu(e_1+...+e_k) = \mu(e_1)+...+\mu(e_k)$ for orthogonal elements $e_1,...,e_k$ in $E$. When $E$ is orthogonally $\sigma$-complete, the state $\mu$ is called $\sigma$-additive if $\mu(\Sigma_n e_n) = \Sigma_n \mu(e_n)$ for any orthogonal sequence $e_n$ in $E$, and when $E$ is orthogonally complete, the state $\mu$ is called completely additive if $\mu(\Sigma_{f \in F} f) = \Sigma_{f \in F} \mu(f)$ for any orthogonal subset $F$ of $E$.

Denote by $S_o$ the set of all states on $E$, by $S_\sigma$ the set of σ-additive states on $E$ and by $S_c$ the set of completely additive states on $E$, where it is assumed that $E$ is orthogonally σ-complete and orthogonally complete, respectively, in the latter two cases.

With a state μ and $\mu(e)>0$ for some $e \in E$, another state ν is called a conditional probability of μ under $e$ if $\nu(f) = \mu(f)/\mu(e)$ holds for all $f \in E$ with $f \prec e$. Now let $S$ be either $S_o$ or $S_\sigma$ or $S_c$. We shall consider the following axioms that were introduced in [21].

(UC1) *If $e,f \in E$ and $\mu(e)=\mu(f)$ for all $\mu \in S$, then $e=f$.*
(UC2) *If $e \in E$ and $\mu \in S$ with $\mu(e)>0$, there is one and only one conditional probability of μ under $e$.*

If these axioms are satisfied, $E$ is called an *S-UCP space* ($S = S_o$, $S_\sigma$, or $S_c$) - named after the major feature of this mathematical structure which is the existence of the <u>u</u>nique <u>c</u>onditional <u>p</u>robability - and the elements in $E$ are called events. The unique conditional probability of μ under $e$ is denoted by $\mu_e$ and, in analogy with probability theory, we also write $\mu(f|e)$ instead of $\mu_e(f)$ for $f \in E$. The above two axioms imply that there is a state $\mu \in S$ with $\mu(e)=1$ for each event $e \neq 0$, that the difference $d$ in (OS6) becomes unique, that the relation $\prec$ is anti-symmetric (but not necessarily transitive), and that $e \perp e$ iff $e \perp \mathbb{I}$ iff $e=0$ ($e \in E$).

Note that the following identity which will be used later holds for convex combinations of states $\mu, \nu \in S$ ($0<s<1$):

$$(s\mu+(1-s)\nu)_e = (s\mu(e)\mu_e+(1-s)\nu(e)\nu_e)/(s\mu(e)+(1-s)\nu(e)). \qquad (1)$$

Examples of the above structure can be obtained considering Jordan algebras. The multiplication operation in a Jordan algebra $A$ satisfies the condition $a^2 \circ (a \circ b) = a \circ (a^2 \circ b)$ for $a,b \in A$. A JB algebra is a complete normed real Jordan algebra $A$ satisfying $\|a \circ b\| \leq \|a\| \|b\|$, $\|a^2\|=\|a\|^2$ and $\|a^2\| \leq \|a^2+b^2\|$ for $a,b \in A$. A partial order relation $\leq$ on $A$ can then be derived by defining its positive cone as $\{a^2 : a \in A\}$. If $A$ is unital, we denote the identity by $\mathbb{I}$. A JB Algebra $A$ that owns a predual $A_*$ (i.e., $A$ is the dual space of $A_*$) is called a JBW algebra and is always unital. A JBW algebra can also be characterized as a JB algebra where each bounded monotone increasing net has a supremum in $A$ and a normal positive linear functional not vanishing in $a$ exists for each $a \neq 0$ in $A$ (i.e., the normal positive linear functionals are separating). A map is normal if it commutes with the supremum. It then turns out that the normal functionals coincide with the predual. The self-adjoint part of any W*-algebra (von Neumann algebra) equipped with the Jordan product $a \circ b := (ab+ba)/2$ is a JBW algebra, but not each JBW algebra is the complete self-adjoint part of a W*-algebra. Moreover, there are exceptional JBW algebras that cannot be represented as an algebra of self-adjoint operators at all (e.g., the algebra of hermitean 3×3 matrices over the octonions equipped with the Jordan product).

The monographs [14] and [26] are recommended as excellent references for the theory of JB/JBW algebras and W*-algebras, respectively. Historically important references are Jordan, von Neumann and Wigner's work [16] which initiated the theory of Jordan algebras and covers the finite-dimensional case, as well as Alfsen, Shultz and Størmer's achievement for the infinite-dimensional case [5].

The idempotent elements of a JB algebra are called projections, and the projections in a JBW algebra $A$ form a complete lattice. These projection lattices now become examples of $S_c$-UCP spaces if $A$ does not contain a type $I_2$ part. The conditional probability has the shape $\mu(f|e)=\hat{\mu}(\{e,f,e\})/\mu(e)$, where $\{a,b,c\} := a \circ (b \circ c) - b \circ (c \circ a) + c \circ (a \circ b)$ is the Jordan triple product, and in the case of a W*-algebra we get $\mu(f|e)=\hat{\mu}(efe)/\mu(e)$. This was shown in [21] and reveals the link to the Lüders - von Neumann quantum measurement. Note that $\hat{\mu}$ on $A$ is the unique linear extension of the state $\mu$ on the projection lattice; this extension exists by Gleason's theorem [13] and its later enhancements to W*-algebras and JBW algebras [8,9,11,20,30,31]. A complete characterization of the projection lattices in JBW algebras without type $I_2$ part as $S_c$-UCP spaces with some further properties will be a major result at the end of this paper.

### 3. Order-unit spaces

A partially ordered real vector space $A$ is an order-unit space if $A$ contains an order-unit $\mathbb{1}$ and if $A$ is Archimedean [1,6,14]. The order-unit $\mathbb{1}$ is positive and, for all $a \in A$, there is $t>0$ such that $-t\mathbb{1} \leq a \leq t\mathbb{1}$. $A$ is Archimedean if $na \leq \mathbb{1}$ for all $n \in \mathbb{N}$ implies $a \leq 0$. An order-unit space $A$ has a norm given by $\|a\| = \inf\{t>0: -t\mathbb{1} \leq a \leq t\mathbb{1}\}$. Each $x \in A$ can be written as $x = a - b$ with positive $a, b \in A$ (e.g., choose $a = \|x\|\mathbb{1}$ and $b = \|x\|\mathbb{1} - x$). A positive linear functional $\rho: A \to \mathbb{R}$ on an order-unit space $A$ is norm continuous with $\|\rho\| = \rho(\mathbb{1})$ and, vice versa, a norm continuous linear functional $\rho$ with $\|\rho\| = \rho(\mathbb{1})$ is positive. Note that unital JB algebras are order-unit spaces.

The order-unit spaces $A$ considered in the following are dual spaces of base-norm spaces $V$ such that the unit ball of $A$ is compact in the weak-*-topology $\sigma(A,V)$. For $\rho \in V$ and $x \in A$ define $\hat{\rho}(x) := x(\rho)$; the map $\rho \to \hat{\rho}$ is the canonical embedding of $V$ in its second dual $V^{**} = A^*$. Then $\rho \in V$ is positive iff $\hat{\rho}$ is positive on $A$. Moreover, $A$ is monotone complete and $\hat{\rho}(\sup x_\alpha) = \lim \hat{\rho}(x_\alpha)$ holds for $\rho \in V$ and any bounded monotone increasing net $x_\alpha$ in $A$; in the JBW/W*-algebra setting one would say that $\rho \in V$ is normal.

For any set $K$ in $A$, denote by $\overline{\text{lin}}\,K$ the $\sigma(A,V)$-closed linear hull of $K$ and by $\overline{\text{conv}}\,K$ the $\sigma(A,V)$-closed convex hull of $K$. For a convex set $K$, denote by $\text{ext}\,K$ the set of its extreme points which may be empty unless $K$ is compact. A projection is a linear map $U: A \to A$ with $U^2 = U$ and, for $a \leq b$, we define $[a,b] := \{x \in A: a \leq x \leq b\}$. Suppose that $E$ is a subset of $[0,\mathbb{1}]$ in $A$ such that

(a) $\mathbb{1} \in E$,
(b) $\mathbb{1} - e \in E$ if $e \in E$, and
(c) $d+e+f \in E$ if $d,e,f,d+e,d+f,e+f \in E$.

Define $e' := \mathbb{1} - e$ and call $e, f \in E$ orthogonal if $e+f \in E$. Then $E$ satisfies (OS1),...,(OS6) such that we can consider $S_o$ as in section 2. Since $[0,\mathbb{1}]$ is monotone complete, any sum of orthogonal elements in $E$ $\sigma(A,V)$-converges in $[0,\mathbb{1}]$. If these sums converge in $E$, we call $E$ orthogonally complete, and we call $E$ orthogonally $\sigma$-complete, if only the countable sums converge in $E$. In these cases, we can again consider $S_\sigma$ and $S_c$ as in section 2.

**Proposition 3.1:** *Suppose that $A$ is an order-unit space with order unit $\mathbb{1}$ and that $A$ is the dual of the base-norm space $V$. Moreover, suppose that $E$ is a subset of $[0,\mathbb{1}]$ satisfying the three above conditions* (a), (b), (c). *Consider the cases $S=S_o$, $S=S_\sigma$, or $S=S_c$, assuming that $E$ is orthogonally $\sigma$-complete or orthogonally complete, respectively, in the latter two cases, and suppose that the following two conditions hold*:

(i) $A = \overline{\text{lin}}\,E$, *and for each $\mu \in S$ there is a $\sigma(A,V)$-continuous positive linear functional $\hat{\mu}$ on $A$ with $\hat{\mu}(e) = \mu(e)$ for $e \in E$.*

(ii) *For each $e \in E$ there is a $\sigma(A,V)$-continuous positive projection $U_e: A \to A$ such that $U_e\mathbb{1} = e$, $U_e A = \overline{\text{lin}}\,\{f \in E: f \leq e\}$ and $\hat{\mu} = \hat{\mu} U_e$ for $\mu \in S$ with $\mu(e) = 1$.*

*Then E is an S-UCP space. The conditional probabilities have the shape* $\mu(f|e) = \hat{\mu}(U_e f)/\mu(e)$ *for* $e, f \in E$ *and* $\mu \in S$ *with* $\mu(e) > 0$.

*Proof.* For $e, f \in E$ with $e \neq f$ there is $\rho \in V_+$ with $\rho(e-f) \neq 0$. The restriction of $\rho/\rho(\mathbb{I})$ to $E$ then yields a state $\mu \in S$ with $\mu(e) \neq \mu(f)$. Therefore (UC1) holds.

Suppose $e \in E$ and $\mu \in S$ with $\mu(e) > 0$. It is rather obvious that the map $g \to \hat{\mu}(U_e g)/\mu(e)$ on $E$ provides a conditional probability of $\mu$ under $e$. Now assume that $\nu$ is a further conditional probability of $\mu$ under $e$. Then $\nu(e) = 1$ and thus $\hat{\nu} = \hat{\nu} U_e$. From $U_e g \in \overline{\text{lin}} \{f \in E: f \leq e\}$ we get that $\nu(g) = \hat{\nu}(U_e g) = \hat{\mu}(U_e g)/\mu(e)$ for $g \in E$. Therefore, (UC2) holds as well.    q.e.d.

Note the similarities between the second part of condition (i) in the preceding proposition and the Gleason theorem. We shall now see that the situation of Proposition 3.1 is universal for the *S*-UCP spaces; i.e., each *S*-UCP space has such a shape as described there.

**Theorem 3.2:** *Suppose that E is a S-UCP space with* $S = S_o$, $S = S_\sigma$, *or* $S = S_c$. *Then E is a subset of* $[0, \mathbb{I}]$ *in some order-unit space A with predual V as described in Proposition* 3.1.

*Proof.* Define $V := \{s\mu - t\nu : \mu, \nu \in S, 0 \leq s, t \in \mathbb{R}\}$, which is a linear subspace of the orthogonally additive real-valued functions on $E$, and consider for $\rho \in V$ the norm $\|\rho\| := \inf\{r \in \mathbb{R} : r \geq 0 \text{ and } \rho \in r \text{ conv}(S \cup -S)\}$. Then $|\rho(e)| \leq \|\rho\|$ for every $e \in E$. Let $A$ be the dual space of the base-norm space $V$ and let $\hat{\mu}$ be the canonical embedding of $\mu \in V$ in $V^{**} = A^*$. If $\hat{\mu}(x) \geq 0$ for all $\mu \in S$, the element $x \in A$ is called positive and we write $x \geq 0$. Equipped with this partial ordering, $A$ becomes an order-unit space with the order-unit $\mathbb{I} := \pi(\mathbb{I})$, and the order-unit norm coincides with the dual space norm such that $\sup\{|\hat{\mu}(x)| : \mu \in S\} = \|x\|$ for $x \in A$. With $e \in E$ define $\pi(e)$ in $A$ via $\pi(e)(\rho) := \rho(e)$ for $\rho \in V$. Then $\|\pi(e)\| \leq 1$, and the finite additivity, $\sigma$-additivity, or complete additivity of $\pi$ in the different cases follow immediately from this definition. Moreover, $A$ is the $\sigma(A, V)$-closed linear hull of $\pi(E)$.

We now define $U_e$ for $e \in E$. Suppose $x \in A$ and $s\mu - t\nu \in V$ with $\mu, \nu \in S$ and $0 \leq s, t \in \mathbb{R}$. Then define $(U_e x)(s\mu - t\nu) := s\mu(e)\hat{\mu}_e(x) - t\nu(e)\hat{\nu}_e(x)$. Here, $\hat{\mu}_e$ and $\hat{\nu}_e$ are the canonical embeddings of the conditional probabilities $\mu_e$ and $\nu_e$ in $A^*$; they do not exist in the cases $\mu(e) = 0$ or $\nu(e) = 0$ and then define $\mu(e)\hat{\mu}_e(x) := 0$ and $\nu(e)\hat{\nu}_e(x) := 0$, respectively. We still have to show that $U_e$ is well defined for $s\mu - t\nu = s'\mu' - t'\nu'$ with $\mu, \mu', \nu, \nu' \in S$ and $0 \leq s, s', t, t'$. Then $s - t = (s\mu - t\nu)(\mathbb{I}) = (s'\mu' - t'\nu')(\mathbb{I}) = s' - t'$ and $s + t' = s' + t$. If $s + t' = 0$, $s = s' = t = t' = 0$ and $U_e x$ is well-defined. If $s + t' > 0$, then either $s\mu(e) + t'\nu'(e) = s'\mu'(e) + t\nu(e) = 0$ and $s\mu(e) = t'\nu'(e) = s'\mu'(e) = t\nu(e) = 0$, or $(s\mu + t'\nu')/(s + t') = (s'\mu' + t\nu)/(s' + t) \in S$ and, calculating the conditional probability under $e$ for both sides of this identity by using (1), we get $s\mu(e)\mu_e + t'\nu'(e)\nu'_e = s'\mu'(e)\mu'_e + t\nu(e)\nu_e$. In all cases, $U_e$ is well defined.

If $\mu(e) = 1$ for $\mu \in S$, then $\mu = \mu_e$ and $\hat{\mu}(U_e x) = (U_e x)(\mu) = \hat{\mu}(x)$ such that $\hat{\mu} = \hat{\mu} U_e$. Thus, $(U_e U_e x)(\mu) = \mu(e)\hat{\mu}_e(U_e x) = \mu(e)\hat{\mu}_e(x) = (U_e x)(\mu)$ for all $\mu \in S$ and hence for all $\rho \in V$ such that $U_e U_e = U_e$, i.e., $U_e$ is a projection. Its positivity, $\sigma(A, V)$-continuity as well as $U_e \mathbb{I} = \pi(e)$ and $U_e \pi(f) = \pi(f)$ for $f \in E$ with $f \leq e$ follow immediately from the definition.

Therefore $\overline{\text{lin}} \{\pi(f): f \in E, f \leq e\} \subseteq U_e A$. Assume $U_e x \notin \overline{\text{lin}} \{\pi(f): f \in E, f \leq e\}$ for some $x \in A$. By the Hahn-Banach theorem, there is $\rho \in V$ with $\hat{\rho}(U_e x) \neq 0$ and $\rho(f) = 0$ for $f \in E$ with $f \leq e$. Suppose $\rho = s\mu - t\nu$ with $\mu, \nu \in S$ and $0 \leq s, t \in \mathbb{R}$. Then $s\mu(f) = t\nu(f)$ for $f \in E$ with $f \leq e$ and thus $s\mu(e)\mu_e(f) = t\nu(e)\nu_e(f)$. The uniqueness of the conditional probability implies $s\mu(e)\mu_e(f) = t\nu(f)\nu_e(f)$, i.e., $\hat{\rho}(U_e f)=0$ for all $f \in E$ such that $\hat{\rho} U_e=0$ which contradicts $\hat{\rho}(U_e x) \neq 0$. This completes the proof of Theorem 3.2 after identifying $\pi(E)$ with $E$.                                                                  q.e.d.

**Lemma 3.3:** *If $e \leq f$ with $e,f \in E$, then $U_e f = e = U_f e$ and $U_e U_f = U_f U_e = U_e$. If $e,f \in E$ are orthogonal, then $U_e f = 0 = U_f e$ and $U_e U_f = U_f U_e = 0$.*

*Proof.* Suppose $e \leq f$. Then $\hat{\mu} U_e U_f = \mu(e) \hat{\mu}_e U_f = \mu(e) \hat{\mu}_e = \hat{\mu} U_e$ for $\mu \in S$, where we have used that $1 = \mu_e(e) \leq \mu_e(f) \leq 1$ for the conditional probability $\mu_e$ implies $\mu_e(f)=1$ and hence $\hat{\mu}_e U_f = \hat{\mu}_e$. Thus $U_e U_f = U_e$. The identity $U_f U_e = U_e$ immediately follows from (ii) in Proposition 3.1. Moreover $e = U_e \mathbb{1} = U_e U_f \mathbb{1} = U_e f$ and $e = U_e \mathbb{1} = U_f U_e \mathbb{1} = U_f e$.

Now assume that $e$ and $f$ are orthogonal. Then $e \leq f'$ and $e = U_e f' = U_e(\mathbb{1}-f) = e - U_e f$ such that $U_e U_f = 0$. In the same way it follows that $U_f U_e = 0$. Therefore $U_f$ vanishes on $U_e A$ such that $U_f U_e = 0$. The identity $U_e U_f = 0$ follows in the same way.                                                      q.e.d.

In the following two sections, it shall be investigated when we have spectral duality in the meaning of Alfsen and Shultz [2] and when $A$ becomes a JBW algebra. This requires the notion of an observable.

## 4. Observables and spectral duality

A bounded real observable $X$ is a bounded spectral measure allocating $e_B \in E$ to each Borel set $B$ in $\mathbb{R}$. Since a spectral measure is $\sigma$-additive, we must assume now that $E$ is an an $S$-UCP space with $S=S_\sigma$ or $S=S_c$. This definition of an observable can be found e.g., in [22,25,28,29]. Some authors use other definitions; however, note that only bounded real-valued sharp observables are considered in the present paper.

The spectral radius of $X$ is $r(X) := \inf\{t \geq 0 : e_{[-t,t]} = \mathbb{1}\}$ and is finite for bounded observables. The probability measure $\mu^X$ with $\mu^X(B) := \mu(e_B)$ is the distribution of $X$ in the state $\mu$, and the measure integral $\int t \, d\mu^X$ is the expectation value of the observable $X$. Then there is a unique element $x$ in $A$ such that $\int t \, d\mu^X = \hat{\mu}(x)$ for all $\mu \in S$ because the map $V \ni \rho \to \int t \, d\rho^X$ yields an element $x$ in $V^* = A$ with $\|x\| = r(X)$. Note that each $e \in E$ and each sum $\Sigma t_k e_k$ with orthogonal events $e_1,...,e_n$ in $E$ and real numbers $t_1,...,t_n$ can be represented by an observable in this way. Such a linear combination of orthogonal events is called a primitive element in $A$. Each $x \in A$ that represents an observable can be approximated by a norm convergent sequence of primitive elements. In general, however, not each element in $A$ represents an observable, which is an important difference to the Jordan algebras of self-adjoint operators usually considered in quantum mechanics.

We are now in the position to formulate some further potential postulates for a system of quantum events. The numbering of the axioms shall remain consistent with the numbering in earlier papers [22,23]; this is the reason why we now come to (A2), (A3) and (A4) while (A1) will be considered in section 5.

(A2)   *For $e,f \in E$ there is a real observable $X$ such that $\mu(f|e)\mu(e) = \int t \, d\mu^X$ for all $\mu \in S$ with $\mu(e)>0$.*

(A3)   *If $Y$ and $Z$ are bounded real observables, there is another real observable $X$ such that $\int t \, d\mu^Y + \int t \, d\mu^Z = \int t \, d\mu^X$ for all $\mu \in S$.*

The first axiom means that the element $U_e f$ in $A$ represents an observable, but is formulated in a more basic way using only the notions of conditional probabilities and observables. The second one postulates the existence of a reasonable sum within the class of real bounded observables. These two axioms were already considered in [22,23], but in a stronger form; there the uniqueness of the observable $X$ was postulated, which we can dispense with in the present paper. For two real observables, $X \leq Y$ shall mean that $\int t \, d\mu^X \leq \int t \, d\mu^Y$ for all $\mu \in S$. This is equivalent to $x \leq y$ for the elements in $A$ that represent the observables.

(A4σ)　　*If $Y_n$ is a bounded (i.e., $r(Y_n) \leq s$ for some $s$) monotone increasing sequence of bounded real observables, there is a real observable X with*
$$\lim_n \int t \, d\mu^{Y_n} = \int t \, d\mu^X$$
　　　　*for all $\mu \in S$.*

(A4c)　　*the same as* (A4σ), *but with bounded monotone increasing nets instead of sequences.*

We shall now study the relationship between the UCP spaces considered here and Alfsen and Shultz's spectral duality [2]. They introduced the so-called P-projections and called $P(\mathbb{I})$ with a P-projections $P$ a projective unit. The $U_e$ considered here are similar to the P-projections, but the two concepts are not identical. A P-projection $P$ has a quasicomplement $Q$ such that $Px=x$ iff $Qx=0$ (and $Qx=x$ iff $Px=0$) for $x \geq 0$. If $U_e x=x$, then $U_{e'}x=0$ by Lemma 3.3, but $U_{e'}x=0$ does not imply $U_e x=x$.

In the case of spectral duality of $V$ and $A$, the system of projective units satisfies (UC1), (UC2), (A2), (A3) and (A4σ) if and only if the states on the projective units have linear extensions to $A$ (as with the Gleason theorem, or in condition (i) of Proposition 3.1). Vice versa, however, what is necessary to get spectral duality from a UCP space? The answer is given in the following proposition.

**Proposition 4.1:** *Suppose that E is a $S_\sigma$-UCP space satisfying* (A2), (A3) *and* (A4σ). *Let L consist of those elements in A that represent observables. Note that L is a monotone σ-complete linear space by* (A3) *and* (A4σ). *Then the following conditions are equivalent*:
(i) *$U_e$ and $U_{e'}$ are quasicomplementary P-projections on L for each $e \in E$ (i.e., L and V are in weak spectral duality).*
(ii) *If $\mu(f)=1$ holds for each state $\mu \in S_\sigma$ with $\mu(e)=1$, then $e \leq f$ ($e,f \in E$).*
*In both cases, $E \subseteq \text{ext}\{x \in L: 0 \leq x \leq \mathbb{I}\}$.*

*Proof.* Assume (i) and that $\mu(f)=1$ holds for each state $\mu \in S_\sigma$ with $\mu(e)=1$ ($e,f \in E$). This means $\mu(e)=\mu(e)\mu_e(f)=\hat{\mu}(U_e f)$ for $\mu \in S_\sigma$; thus $e=U_e f$ and $0=U_e f'$. Since $U_e$ and $U_{e'}$ are quasicomplementary P-projections, we have for positive $x$ in $L$ that $U_e x=0$ iff $U_{e'}x=x$. Therefore, $e' \geq U_{e'} f' = f'$ and $e \leq f$.

Now suppose (ii). This means that $0=U_e f'$ implies $e \leq f$ for any $e,f \in E$. If now $0=U_e x$ for some positive $x$ in $L$, then $0=U_e p$ for any $p$ in a spectral resolution of $x$ such that $e \leq p'$ and $p \leq e'$; therefore $U_{e'} p = p$ for all $p$ in the spectral resolution and finally $U_{e'}x=x$.

Assume $e \in E$ and $e=sx+ty$ with $x,y \in L$, $0 \leq x,y \leq \mathbb{I}$ and real number $s,t>0$, $s+t=1$. Then $0=U_{e'}e=sU_{e'}x+tU_{e'}y$ such that $U_{e'}x=0=U_{e'}y$. The quasicomplementarity yields $x=U_e x \leq e$ and $y=U_e y \leq e$ such that the identity $e=sx+ty$ is possible only if $e=x=y$. Therefore $e$ is an extreme point in $\{x \in L: 0 \leq x \leq \mathbb{I}\}$.　　　　　　　　　　　　　　　　　　　　q.e.d.

Note that Alfsen and Shultz distinguish between spectral duality and weak spectral duality; spectral duality means existence and uniqueness of the spectral resolution, while the weak form does not require the uniqueness [2]. We would achieve spectral duality in condition (i) in the above proposition, when we would assume uniqueness for the observables in (A2) (A3) and (A4σ).

For the results in the following section, it will neither be necessary to assume that condition (ii) of Proposition 4.1 holds nor that the $U_e$ are P-projections. This is why the approach of the present paper differs from other approaches that assume spectral duality [2-4,7,15] or at least the existence of sufficiently many P-projections [12].

## 5. Jordan operator algebras

We now arrive at the last missing ingredient to make $A$ a JBW algebra. This is the following axiom.

(A1)　　$\mu(f|e)\mu(e) + \mu(f'|e')\mu(e') = \mu(e|f)\mu(f) + \mu(e'|f')\mu(f')$ *for $e,f \in E$ and $\mu \in S$.*

(A1) is equivalent to $U_e f + U_{e'} f' = U_f e + U_{f'} e'$ for $e,f \in E$. In this form, (A1) occurred for the first time in [4] for the P-projections considered there. Note that (A1) holds in W*-algebras and JBW algebras.

**Theorem 5.1:** *Suppose that $E$ is a S-UCP space with $S=S_\sigma$ or $S=S_c$ and that* (A1), (A2), (A3) *hold. Then $A$ can be equipped with a* (*generally non-associative*) *commutative multiplication operation ∘ such that $A$ becomes a JBW algebra and $E$ is a $\sigma(A,V)$-dense subset of the projection lattice in $A$; $\mathbb{I} \in E$ becomes the unit element of the JBW algebra.*

*Proof.* Denote by $L$ the subset of $A$ containing all elements which represent observables and by $M$ its norm closure. Because of (A3) $L$ and thus also $M$ are linear subspaces of $A$. Define $T_e x := \frac{1}{2}(x + U_e x - U_{e'} x)$ for $e \in E$ and $x \in A$. Then $T_e x \in L$ for primitive elements $x$ by (A2) and (A3) and thus $T_e x \in M$ for $x \in M$ due to the norm continuity of $U_e$ and $U_{e'}$. Moreover, $U_e[-\mathbb{I}, \mathbb{I}] \subseteq [-e, e]$ and $U_{e'}[-\mathbb{I}, \mathbb{I}] \subseteq [-e', e']$, thus $(U_e - U_{e'})[-\mathbb{I}, \mathbb{I}] \subseteq [-\mathbb{I}, \mathbb{I}]$ and therefore $\|U_e - U_{e'}\| \leq 1$. Hence $\|T_e\| \leq 1$.

(A1) implies $T_e f = T_f e$ for $e,f \in E$. For $x \in M$ and a primitive element $y = \Sigma t_k e_k$ with orthogonal events $e_1,...,e_n$ in $E$ and real numbers $t_1,...,t_n$ define $T_y x := \Sigma t_k T_{e_k} x$ for $x \in A$. Then $T_y x = T_x y$ for primitive elements $x$ and $y$, and this also implies that $T_y$ is well-defined. Furthermore, $\|T_y e\| = \|T_e y\| \leq \|y\|$ for $e \in E$ and thus $\|T_y x\| \leq \|x\| \|y\|$ for all primitive elements $x,y$ since $x$ with $\|x\| \leq 1$ is a convex combination of elements from $E$ and $-E$. Now define $x \circ y$ for $x,y \in M$ as the norm-continuous extension of $T_x y$ to $M$.

Then $e^2 := e \circ e = e$ and $e \circ f = 0$ for orthogonal events $e,f \in E$. For a primitive element $y = \Sigma t_k e_k$ we have $y^n = \Sigma t_k^n e_k$. This implies $\|y^2\| = \max\{t_k^2\} = (\max\{|t_k|\})^2 = \|y\|^2$. Moreover, from $\|x\| = \sup\{|\hat{\mu}(x)| : \mu \in S\}$ we get $\|x^2\| \leq \|x^2 + y^2\|$ for primitive elements $x$ and $y$. Due to norm continuity, $\|y^2\| = \|y\|^2$ and $\|x^2\| \leq \|x^2 + y^2\|$ then hold for $x,y \in M$ and, since the primitive elements are power-associative, each element in $M$ is power-associative. By a result in [16] or [27], this implies $e \circ (f \circ y) = f \circ (e \circ y)$ for orthogonal idempotent elements $e$ and $f$ and for any $y$. Therefore we get for a primitive element $x = \Sigma t_k e_k$ with orthogonal events $e_1,...,e_n$ in $E$ and $y \in M$:

$$x^2 \circ (x \circ y) = \Sigma_k \Sigma_l t_k^2 t_l e_k \circ (e_l \circ y) = \Sigma_k \Sigma_l t_k^2 t_l e_l \circ (e_k \circ y) = x \circ (x^2 \circ y).$$

Due to norm continuity, this identity then holds for all $x,y \in M$ such that $M$ become a JB algebra.

We now consider the seminorms $x \to \hat{\mu}(x^2)^{1/2}$, $\mu \in S$, on $M$ and the topology defined by them on $M$ which is called the *s*-topology. Norm convergence implies *s*-convergence. Due to the Cauchy-Schwarz inequality, *s*-convergence implies $\sigma(M,V)$-convergence, the product $x \circ y$ is *s*-continuous separately in each factor and jointly *s*-continuous in both factors on bounded subsets of $M$. Let $N$ comprise all those elements of $A$ that are the $\sigma(A,V)$-limit of a bounded net in $M$ which is a Cauchy net with respect to the seminorms defining the *s*-topology. Then $\hat{\mu}(x^2) \geq 0$ for $\mu \in S$, $x \in N$, and $N$ becomes a JB algebra since it inherits all the properties from $M$. Note that the $\sigma(A,V)$-closure of $M$ is not automatically a JB algebra since the map $x \to x^2$ is not $\sigma(A,V)$-continuous.

If now $x_\alpha$ is a bounded monotone increasing net in $N$, then $x_\alpha$ $\sigma(A,V)$-converges to $\sup x_\alpha$ in $A$. Since $(x-y)^2 \leq \|x-y\|(x-y)$ for $y \leq x$ in a JB algebra, the net $x_\alpha$ *s*-converges to $\sup x_\alpha$ such that $\sup x_\alpha \in N$. Therefore $N$ is monotone complete and the restrictions of the positive elements of $V$ to $N$ provide a separating family of normal functionals such that $N$ becomes a JBW algebra. On the other hand, if $\rho$ is any positive normal functional on $N$ with $\rho(\mathbb{1})=1$, its restriction to $E$ belongs to $S$ such that $V$ becomes identical with the normal functionals on $N$ or with the predual of $N$. Thus $A = V^* = N$ is a JBW algebra.

Since the unit ball of $M$ is contained in the norm closure of $\text{conv}(E-E)$, the $\sigma(A,V)$-closure of the unit ball of $M$, which is the unit ball of $A$, must be included in the $\sigma(A,V)$-closure $\overline{\text{conv}}(E-E)$. Therefore $[-\mathbb{1},\mathbb{1}]=\overline{\text{conv}}(E-E)$. Applying the $\sigma(A,V)$-continuous affine function $x \to (x+1)/2$ to both sides and using the identity $(e-f+1)/2=(e+f')/2$ we get $[0,\mathbb{1}]=\overline{\text{conv}}(E)$. The Krein-Milman theorem then implies that $\text{ext}[0,\mathbb{1}] \subseteq \overline{E}$. Since, in a JBW algebra, $\text{ext}[0,\mathbb{1}]$ coincides with the projection lattice which contains $E$, we finally have the $\sigma(A,V)$-density of $E$ in $\text{ext}[0,\mathbb{1}]$.      q.e.d.

Theorem 5.1 is an improvement of an earlier result by the author [22] in two ways. Only the existence of the observable $X$ must be assumed in (A2) and (A3), the uniqueness postulate becomes redundant. Furthermore, in Theorem 5.1, the embedding of $E$ in the JBW algebra is $\sigma$-additive and completely additive in the different cases due to Theorem 3.2, while the one considered in [22] is only finitely additive.

**Corollary 5.2:** *Suppose that $E$ is a $S_c$-UCP space and that* (A1), (A2), (A3), (A4c) *hold. Then $E$ is identical with the projection lattice of a JBW algebra.*

*Proof.* Consider the subset $L$ of $A$ containing all elements which represent observables. Due to (A4c) $L$ is monotone complete and thus norm complete since sequential monotone completeness already implies norm completeness (e.g., see [2] or [25]). In the proof of Theorem 5.1 we have seen that $L$ is a JB algebra then. The positive part of $V$ provides a separating family of normal functionals. Thus $L$ is a JBW algebra (and $L=N=A$). Since each $x$ in $L$ represents an observable, its spectral resolution belongs to $E$; applying this to a projection $x=p$ yields $p \in E$.      q.e.d.

**Corollary 5.3:** *Suppose that $E$ is a $S_\sigma$-UCP space and that* (A1), (A2), (A3), (A4$\sigma$) *hold. Then $E$ is identical with the projection lattice of a monotone sequentially complete JB algebra with a separating family of $\sigma$-normal states.*

This follows in the same way as the preceding corollary. The monotone sequentially complete JB algebras appearing in Corollary 5.3 are by far not as well-known as the JBW algebras, but are nevertheless considered here because they appear a more natural analogue of the measurable function spaces of probability theory than the JBW algebras; indeed, the measurable functions form a monotone sequentially complete JB algebra, but not a JBW algebra in general. The monotone sequentially complete C*-algebras were studied by Christensen [10], Kadison [18], Kehlet [19] and Pedersen [24].

Corollary 5.2 provides a complete characterization of the projection lattices in JBW algebras without type $I_2$ part similar to the one provided by Bunce and Wright [7]. However, they use other characterizing properties; besides the countable chain condition and an "elliptic" state space they assume spectral duality in the meaning of Alfsen and Shultz [2] which presumes the existence of sufficiently many P-projections and the uniqueness for the spectral measure. They do not use (A1) which is replaced by the "ellipticity". Their result is based on other characterizations that do not concern the projection lattice, but the state space of an operator algebra [2-4,15]. These characterizations involve very interesting geometric properties of the state space and are extremely satisfying from a mathematical viewpoint, but they become less satisfactory if one's concern is the axiomatic foundation of quantum mechanics because plausible interpretations and physical meanings are hard to find for these properties.

## 6. Conclusions

In classical probability theory, the observables or random variables are modeled by measurable functions. When proceeding to quantum mechanics, self-adjoint operators must be used as model instead. With the still more general framework of the present paper, the observables become elements of an order-unit space $A$ like in Proposition 3.1, but not each $a \in A$ represents an observable. This is a significant difference to standard quantum mechanics (and even to the JBW algebra model). Moreover, the observables cannot anymore be represented as self-adjoint linear operators on a Hilbert space such that discussions about the so-called "collapse of the wave-function" and its interpretation would become needless.

A concrete example of a UCP space that does not have a Hilbert space representation is given by the projection lattice in the algebra of hermitean 3×3 matrices over the octonions which is an exceptional Jordan algebra. An example of a UCP space where $A$ is not a JBW algebra is not known at present.

Although the axioms used in the present approach and in the characterization of the projection lattices in section 5 have plausible interpretations and physical meanings, it is still not clear whether each single one is an absolute must in an axiomatic foundation of quantum mechanics. The axiom (A1) turns out to be very useful, but does not appear to be a really natural one. The postulate (A3) concerning the existence of a reasonable sum within the class of real observables is usually considered a quite natural axiom, but wouldn't it then be as natural for the vector-valued observables where it is already violated by standard quantum mechanics and disproved by physical experiments?

In the present paper, two extremes have been considered - the UCP spaces with only the basic properties and those with so much structure that they become projection lattices in JBW algebras. There may be further interesting structures between the two extremes satisfying other combinations of the axioms; the combination (UC1) and (UC2) together with (A1) appears interesting in so far as these three postulates require only the basic notion of conditional probability and do not need the notion of an observable.

Alfsen and Shultz's spectral duality of an order-unit space and a base-norm space [2] has been identified as a candidate of a mathematical structure lying between the two extremes; in this case, (A1) does not hold, but all other axioms would do, if the Gleason extension theorem were satisfied for the states on the projective units. However, all the concrete examples considered by Alfsen and Shultz either are related to JBW algebras, or they contain at most two orthogonal non-zero events (like the type $I_2$ JBW factors) such that they cannot satisfy the Gleason extension theorem unless being identical with the Boolean algebra $\{0,e,e',\mathbb{I}\}$. Another candidate structure are the GL spaces considered by Edwards and Rüttimann [12]; in this case, only (UC1) and (UC2) would hold, if the Gleason extension theorem were available, and the $U_e$ become P-projections.